\documentclass[journal,twoside]{IEEEtran}
\usepackage[OT1]{fontenc}

\usepackage{amsmath}
\usepackage{amsthm}
\usepackage{amssymb}
\usepackage[linesnumbered,lined,ruled,vlined]{algorithm2e}
\usepackage{bm}
\usepackage{booktabs}
\usepackage{cite}
\usepackage{cases}
\usepackage[usenames, dvipsnames]{color}
\usepackage{colortbl}
\usepackage{float}
\usepackage[colorlinks, urlcolor=blue, linkcolor=black, anchorcolor=black, citecolor=black]{hyperref}
\usepackage{url}
\usepackage{graphicx}
\usepackage{subfigure}
\usepackage{lipsum}
\usepackage{makecell} 
\usepackage{multirow}
\usepackage{microtype} 
\usepackage{mathtools}
\usepackage{subfigure}
\usepackage{tabularx}
\usepackage{threeparttable}  
\usepackage{xfrac}
\usepackage{balance}
\usepackage{xcolor}
\usepackage{stfloats} 
\usepackage{diagbox} 
\usepackage{pgfplots} 

\definecolor{mygray}{gray}{.9}

\theoremstyle{definition}

\definecolor{mygreen}{HTML}{00B050}

\begin{document}
	
	\title{Belief-selective Propagation Detection \\ for MIMO Systems}
	\author{Wenyue Zhou, 
            Yifei Shen, 
            Liping Li, 
            Yongming Huang,~\IEEEmembership{Senior Member,~IEEE},\\
            Chuan Zhang,~\IEEEmembership{Senior Member,~IEEE},
            and Xiaohu You,~\IEEEmembership{Fellow,~IEEE}
\thanks{W. Zhou, Y. Shen, Y. Huang, X. You, and C. Zhang are with the National Mobile Communications Research Laboratory, Southeast University, Nanjing 210096, China, and also with the Purple Mountain Laboratories, Nanjing 211189, China. Email: \{chzhang,xhyu\}@seu.edu.cn. \emph{(Corresponding authors: Chuan Zhang and Xiaohu You.)}}
\thanks{L. Li is with the Key Laboratory of Intelligent Computing and Signal Processing, Ministry of Education, Anhui University, Hefei 230039, China.}
	}
	
	

	\maketitle
	
	\begin{abstract}

Compared to the linear MIMO detectors, the Belief Propagation (BP) detector has shown greater capabilities in achieving near optimal performance and better nature to iteratively cooperate with channel decoders.
Aiming at real applications, recent works mainly fall into the category of reducing the complexity by simplified calculations, at the expense of performance sacrifice. However, the complexity is still unsatisfactory with exponentially increasing complexity or required exponentiation operations. Furthermore, due to the inherent loopy structure, the existing BP detectors persistently encounter error floor in high signal-to-noise ratio (SNR) region, which becomes even worse with calculation approximation.
This work aims at a revised BP detector, named {Belief-selective Propagation (BsP)} detector by selectively utilizing the \emph{trusted} incoming messages with sufficiently large \textit{a priori} probabilities for updates. Two proposed strategies: symbol-based truncation (ST) and edge-based simplification (ES) squeeze the complexity (orders lower than the Original-BP), while greatly relieving the error floor issue over a wide range of antenna and modulation combinations. For the $16$-QAM $8 \times 4$ MIMO system, the $\mathcal{B}(1,1)$ {BsP} detector achieves more than $4$\,dB performance gain (@$\text{BER}=10^{-4}$) with roughly $4$ orders lower complexity than the Original-BP detector. Trade-off between performance and complexity towards different application requirement can be conveniently obtained by configuring the ST and ES parameters.
	\end{abstract}
	
	\begin{IEEEkeywords}
	Multiple-input multiple-output (MIMO),
    Belief-selective Propagation (BsP), error floor, complexity.
	\end{IEEEkeywords}

	\IEEEpeerreviewmaketitle

	\section{Introduction}

\IEEEPARstart{M}{ULTIPLE-INPUT MULTIPLE-OUTPUT (MIMO)} ranks among the key enabling technologies for nowadays wireless systems by offering high spectrum efficiency (SE) and energy efficiency (EE) \cite{Hoydis13_JSAC,Geoffrey14_JSTSP} . Therefore, designing efficient MIMO detection algorithm balancing both performance and complexity becomes one important topic \cite{rusek2012scaling}.
Although the \emph{maximum a posteriori} (MAP) detection or the \emph{maximum likelihood} (ML) detection leads to minimum error probability, the exponentially increasing complexity with modulation order and MIMO scale hinders real applications.
To this end, the sphere decoder (SD) \cite{damen2000lattice} has been proposed to reduce complexity by restricting candidate symbols to ones within a sphere (depth-first) \cite{Brink03_TCOM} or a list of size $K$ (width-first) \cite{Guo06_JSAC}. Providing sub-optimal performance with acceptable complexity, SD enjoys a very successful application for small-scale MIMO. However, for large-scale MIMO, it still suffers from high complexity penalty, which increases in cubic (often roughly) with system scale \cite{hassibi2005sphere}.
For large-scale MIMO, one compromise choice is to employ linear detector such as zero-forcing (ZF) detector or more often the linear minimum mean square error (LMMSE) detector \cite{rusek2012scaling}. Though the linear detector is much simpler and more hardware friendly, its performance is much worse than that of the MAP/ML detector or SD \cite{Yang15_CSTO}. Furthermore, for large-scale MIMO, the excessive complexity resulting from matrix inversion becomes a problem.
By making use of the channel sparsity \cite{narasimhan2014channel}, approaches such as \emph{Neumann} series approximation (NSA) \cite{zhang2018expectation} and iterative methods \cite{wu2016efficient} have been considered to address the matrix inversion issue. However, the performance of these linear variants is still bounded by that of the LMMSE detector, and cannot fully exploit the MIMO benefits.

With the fast development of wireless communications, especially the emergence of new service scenarios, such as the enhanced mobile broadband (eMBB) service and the low-latency communications (URLLC) service \cite{ITU_R_2017,NR_2017}, MIMO detectors which can deliver performance beyond the LMMSE curve with good implementation feasibility are highly expected\footnote{Admittedly, the satisfactory system performance can still be achieved by employing more advanced channel codes with LMMSE detector. This topic is out of the scope of this paper and will be discussed in another paper of ours.}.
Among all candidates, the Belief Propagation (BP) detector \cite{Jun08_JSAC}, which is based on the Bayes' rule \cite{pearl1988probabilistic, Kschischang01_TIT} has recently drawn increasing attentions from both academia and industry.
Although the BP detector, at least for now is not an ``all-around winner'' for different performance/complexity requirements, it has shown considerable potential as a well-balanced solution with the following merits. First, the BP detector can deliver near optimal performance, and outperforms the linear ones. Second, the BP detector is soft-input and soft-output, and can iteratively work with channel decoders and other baseband modules in nature. Third, thanks to the regularity of factor graph (FG) the BP detector is hardware-friendly, and can be easily parallelized and scaled for applications.

The BP MIMO detection with the FG representation is originally proposed in \cite{Jun08_JSAC} for the Bell labs layered space-time (BLAST) architecture \cite{foschini1996layered}, which is denoted as the Original-BP in this paper. It can achieve near optimal performance in small-to-medium scale MIMO, with the cost of exponentially increasing complexity. The following works mainly focus on reducing the complexity while maintaining the performance advantage of BP detector. The approaches can be roughly categorized into two groups: simplifying the edges of the FG, and simplifying the nodes of the FG.

\emph{Edge reduction}: To lower the complexity, the regular-$d_f$ BP (EBRDF-BP) detection is proposed in the same paper \cite{Jun08_JSAC}, which is in fact an edge-pruning approach by remaining the edges corresponding to the $d_f$ largest channel coefficients whereas disconnecting the others. Simulation results have shown the complexity reduction with acceptable performance degradation.


\emph{Node reduction}: A major part of related literatures work on the node reduction. In \cite{som2011low}, the BP detection with the \emph{Gaussian} approximation of interference (GAI-BP) has been proposed to lower the node calculation. In \cite{Feichi11_SS}, the single edge-based BP detection with the pseudo \emph{priori} (PP) information initialization and the \emph{Gaussian} approximation feedback (GF) information has been proposed. This PP-GF-BP algorithm further lower the detection complexity. Since both the GAI-BP and the PP-GF-BP detectors are bit-based, their performance with high-order modulations has been sacrificed. In \cite{yang2018low}, the GAI-BP detector is extended to the real domain (RD-GAI-BP), which successfully outperforms both the GAI-BP and the PP-GF-BP detectors with high-order modulations.

So far, the design of efficient BP MIMO detectors remaining challenging. First, the existing BP detectors including the Original-BP detector suffer an aggravate performance error floor in relatively high signal-to-noise ratio (SNR) region. It is caused by both approximations such as GAI and the inherent loopy structure of the full-connected FG model. Second, the complexity of existing BP detectors is still high and not flexible enough for various applications. These multiple challenges motivate us to think of an approach which can reduce the detection complexity while modifying the loopy structure towards better performance.


In this paper, {a Belief-selective Propagation (BsP)} detection algorithm is proposed, which only utilizes the incoming messages with relatively large \textit{a priori} probabilities to update the output messages. Avoiding to propagate the messages with ``low beliefs", the BsP detector cannot only enjoy lower complexity, but also get rid of the performance error floor. It is noted that a similar idea of the BsP was previously proposed for decoding non-binary low-density parity-check (NB-LDPC) codes \cite{Declercq05_ISIT,Declercq07_TCOM,Declercq10_TCOM}, as extended min-sum (EMS). The initial motivation of EMS was only for lower complexity but not for better performance. Second, its application for MIMO detection is not straightforward.
Compared to the EMS decoding, the proposed {BsP} detection differs in the message update for factor nodes.
The BsP detection updates the beliefs by exhaustively searching for the most likely transmitted symbol vector in the given set, whereas the EMS decoding directly computes the messages associated with symbols using the check-node constraints.

\textit{{Contributions:}} The contributions of this work are listed as follows:
\begin{itemize}
  \item The BsP MIMO detection has been first proposed. Compared with the state-of-the-art (SOA) BP detections, it can offer better performance with lower complexity. The persistent error-floor issue of existing BP detections, caused by the loopy FG model has been greatly relieved.
  \item The symbol-based truncation (ST) strategy and the edge-based simplification (ES) strategy are proposed to make the BsP detection feasible for applications. With both strategies, the transmitted symbols corresponding to relatively small probabilities are removed from the search space in the node message update. Therefore, the complexity of the BsP detection has been further reduced. Both strategies are equipped with adjustable parameters to flexibly meet different performance/complexity requirements.
  \item The PP information provided by an LMMSE detector is utilized to generate the initial \textit{a priori} information, which is essential to the BsP detection. This initial \textit{a priori} information effectively contributes to the performance advantages of the BsP detection.
\end{itemize}

In addition, numerical results confirm that the advantages of BsP detection over its counterparts hold in various MIMO scenarios. This robustness guarantees its wide applications.

The reminder of this paper is organized as follows. In Section \ref{sec:preliminaries}, the system model and the Original-BP detection algorithm are reviewed. In Section~\ref{sec:EMBsP}, the BsP detection algorithm is proposed with details.
The computational complexity issues of the BsP are analyzed and discussed in Section~\ref{sec:Complexity}. The numerical results are given in Section~\ref{sec:Results}. Finally, Section~\ref{sec:conclusion} concludes the entire paper.

        \section{Preliminaries}\label{sec:preliminaries}

    \subsection{System Model and the Optimal Solution}\label{subsec:system}
Consider a MIMO system with $N_r$ receive antennas and $N_t$ transmit antennas, then the transmitted symbol vector is an $N_t \times 1$ i.i.d. vector $\mathbf{s}=[s_1,s_2,...,s_{N_t}]^T$. Assume all of the transmitted symbols are from the same QAM modulation of order $M$, then the cardinality of the constellation is $\mathcal{|A|}=2^M$.
Let the $N_r \times N_t$ matrix $\mathbf{H}=[\mathbf{h}_1^T,\mathbf{h}_2^T,...,\mathbf{h}_{N_r}^T]^T$ denote the flat-fading complex MIMO channel, where $\mathbf{h}_i$ represents the $i$-th row vector of $\mathbf{H}$, and each component of $\mathbf{h}_i$ is a complex channel coefficient following the zero-mean and unit-variance Gaussian distribution.
The received signal $\mathbf{y}=[y_1,y_2,...,y_{N_r}]^T$ is given by
\begin{equation}\label{eq:model}
\mathbf{y} = \mathbf{Hs} + \mathbf{n},
\end{equation}
where $\mathbf{n}=[n_1,n_2,...,n_{N_r}]^T$ is a noise vector.
The components of $\mathbf{n}$ are complex independent white Gaussian variables with zero mean and $\sigma^2$ variance.
	
The goal of the MIMO detection is to recover the transmitted symbol vector $\mathbf{s}$ according to the received signal $\mathbf{y}$ and the channel matrix $\mathbf{H}$. The optimal solution to achieve this goal is the MAP detection, which requires an exhaustive search to find the transmitted symbol vector. This MAP detection is illustrated by
\begin{equation}\label{eq:MAP}
\hat{\mathbf{s}}=\mathop{\mathrm{arg}\ \mathrm{max}}_{\mathbf{s}\in\mathcal{A}^{N_t}}\ P(\mathbf{s}|\mathbf{y},\mathbf{H}),
\end{equation}
where the \textit{a posteriori} probability $P(\mathbf{s}|\mathbf{y},\mathbf{H})$ can be rewritten as $P(\mathbf{y}|\mathbf{s},\mathbf{H})P(\mathbf{s})$ employing the Bayes' rule.
Although the MAP detection has the optimal error rate performance, it suffers from an exponentially increasing computational complexity dominated by the number of transmit antennas and the order of the modulations.

    \subsection{BP Detection with the FG Model}\label{subsec:BP}
\begin{figure}[t]
  \centering
  \includegraphics[width=0.7\linewidth]{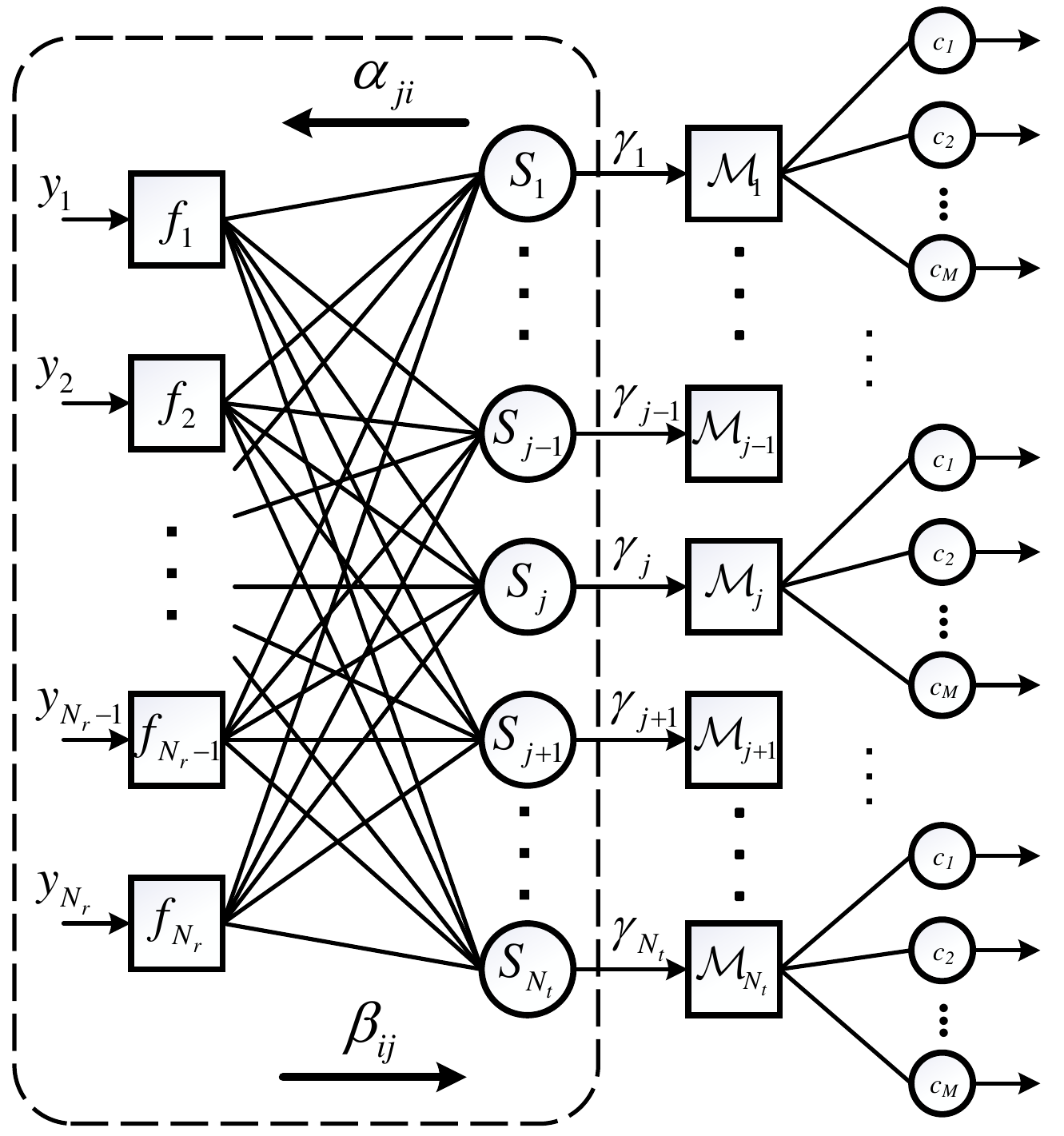}
  \caption{FG model of the MIMO BP detection.}
  \label{fig:factorGraph}
\end{figure}

Apart from the optimal solution, the BP algorithm based on the FG model is an alternative solution for the MIMO detection.
Generally, there are two types of nodes in a FG model, namely factor nodes $\{f_i\}$ and symbol nodes $\{S_j\}$. Fig.~\ref{fig:factorGraph} shows the FG model of the BP detection. This model is related to the channel matrix $\mathbf{H}$.
As shown in Fig.~\ref{fig:factorGraph}, the factor nodes $(f_1, f_2, ..., f_{N_r})$ associate with the received signals $(y_1,y_2,...,y_{N_r})$, and the symbol nodes $(S_1,S_2,...,S_{N_t})$ link with the symbol soft messages $(\gamma_1,\gamma_2,...,\gamma_{N_t})$. Each factor node $f_i$ corresponds to the $i$-th row of the channel matrix, and each symbol node $S_j$ relates to the $j$-th column of the channel matrix.
Every pair of factor node $f_i$ and symbol node $S_j$ are connected with an edge corresponding to the channel coefficient $h_{ij}$.

With the received signal $\mathbf{y}$, the BP detector begins with the update of the $\beta$ messages.
These $\beta$ messages will be delivered from the factor nodes to the symbol nodes.
Then the BP detector computes the $\alpha$ messages with the incoming $\beta$ messages for each symbol node, and transmits them in a reverse direction.
The BP detector iteratively updates the $\beta$ and $\alpha$ messages, until it achieves the maximum number of the iterations.
The BP detection ends up with computing the output soft messages of the coded bits as shown in Fig.~\ref{fig:factorGraph}.

In this paper, log-likelihood ratios (LLRs) are used to represent the soft messages.
The $\alpha$ message passing from the $j$-th symbol node to the $i$-th factor node at the $l$-th iteration is denoted as
\begin{equation}\label{eq:LLR}
\alpha_{ji}^{(l)}(k) = \mathrm{log}\frac{p^{(l)}(s_j=\mu_k)}{p^{(l)}(s_j=\mu_1)},
\end{equation}
where $\mu_k$ ($k\in \{1,2,...,|\mathcal{A}|\}$) denotes the $k$-th modulation symbol in the constellation.
The details of the message update in the BP detection are as follows.

\textit{1) $\beta$ Message Update:}
To update the $\beta$ messages, the factor nodes receive the $\alpha$ messages in the preceding iteration from the symbol nodes, then operate with the following formula
\begin{small}
\begin{equation}\label{eq:simpleBeta}
\begin{aligned}
\beta_{ij}^{(l)}(k) =  & \mathop{\mathrm{max}}_{\mathbf{s}:s_j=\mu_k}
\left \{-\frac{1}{2\sigma^2}\|y_i - \mathbf{h}_i \mathbf{s}\|^2 + \mathop{\sum}_{t=1,t\neq j}^{N_t} \alpha_{ti}^{(l-1)} \right \} \\
 & - \mathop{\mathrm{max}}_{\mathbf{s}:s_j=\mu_1}  \left \{-\frac{1}{2\sigma^2}\|y_i - \mathbf{h}_i \mathbf{s}\|^2 + \mathop{\sum}_{t=1,t\neq j}^{N_t} \alpha_{ti}^{(l-1)} \right \},
\end{aligned}
\end{equation}
\end{small}
where $\beta_{ij}^{(l)}$ denotes the $\beta$ message passing from the $i$-th factor node to the $j$-th symbol node at the $l$-th iteration.
The detailed derivation of (\ref{eq:simpleBeta}) can be referred to \cite{Jun08_JSAC}.

\textit{2) $\alpha$ Message Update:}
In the $l$-th iteration, the symbol nodes compute the $\alpha$ messages by
\begin{equation}\label{eq:alphaMessage}
\alpha_{ji}^{(l)}(k) = \mathop{\sum}_{t=1, t\neq i}^{N_r} \beta_{tj}^{(l)}(k),
\end{equation}
which only involves the addition operations. Therefore, the $\alpha$ message update has much less computational complexity compared with the $\beta$ message update.

The BP detector terminates the message update when reaching the maximum iteration number, and then computes the symbol LLRs by
\begin{equation}\label{eq:output}
\gamma_{j}(k) = \sum \limits_{i=1}^{N_r} \beta_{ij}(k).
\end{equation}
Given a symbol LLRs, its associated coded bit LLRs can be derived according to the FG of the factor nodes $\mathcal{M}_j$ ($j\in \{1,2,...,N_t\}$) and the related bit nodes $c_m$ ($m\in \{1,2,...,M\}$) in Fig.~\ref{fig:factorGraph}. With the sum-product algorithm and the max-log approximation, the coded bit LLRs can be computed by
\begin{equation}\label{eq:newBitLLR}
r_j(m) = \mathop{\mathrm{max}}_{\mu_k:c_m=1}\gamma_j(k)
- \mathop{\mathrm{max}}_{\mu_k:c_m=0}\gamma_j(k),
\end{equation}
where $r_j(m)$ is the $m$-th bit LLR associated with the $j$-th symbol node.
More details about (\ref{eq:newBitLLR}) can be found in \cite{Nassar14_TSP}.

    \section{The Proposed BsP Detection}\label{sec:EMBsP}
The Original-BP detection requires an exhaustive search space involving all of the possible choices of the transmitted symbol vector to update the message $\beta_{ij}$ for each factor node. Such an exhaustive search space suffers from an exponentially increasing size dominated by the constellation cardinality ($|\mathcal{A}|$, as the base) and the transmit antenna number ($N_t$, as the exponent).
Therefore, the Original-BP detection leads to a high computational complexity for MIMO systems.
Fortunately, it is observed that the transmitted symbol vector with relatively low reliability has limited contribution to the accuracy of the updated message $\beta_{ij}$. Therefore, squeezing the search space by removing the low-reliability transmitted symbol vectors is an intuitively reasonable measure to bring down the complexity of the Original-BP detection.
Second, it is expected that the search space reduction will also mitigate the loop issue in the FG by eliminating unnecessary message passing, therefore improve the performance.

Based on this idea, a low-complexity BsP detection algorithm is proposed in this paper.
The BsP detection relies on the following two strategies to shrink the search space:
1) ST strategy to bring down the base of the exponential for the search space size, and 2) ES strategy to reduce the exponent of the exponential for the search space size.
In addition, an LMMSE detector is employed to provide the PP information, resulting in a more efficient initialization for the \textit{a priori} information of the transmitted symbols.
In this section, the ST strategy, the ES strategy and the initialization approach with the LMMSE detector are introduced first, then the proposed BsP detection is summarized as an algorithm.

\begin{figure*}[htbp]
  \centering
  \subfigure{
    \begin{minipage}[t]{0.4\linewidth}
    \centering
    \includegraphics[width=\linewidth]{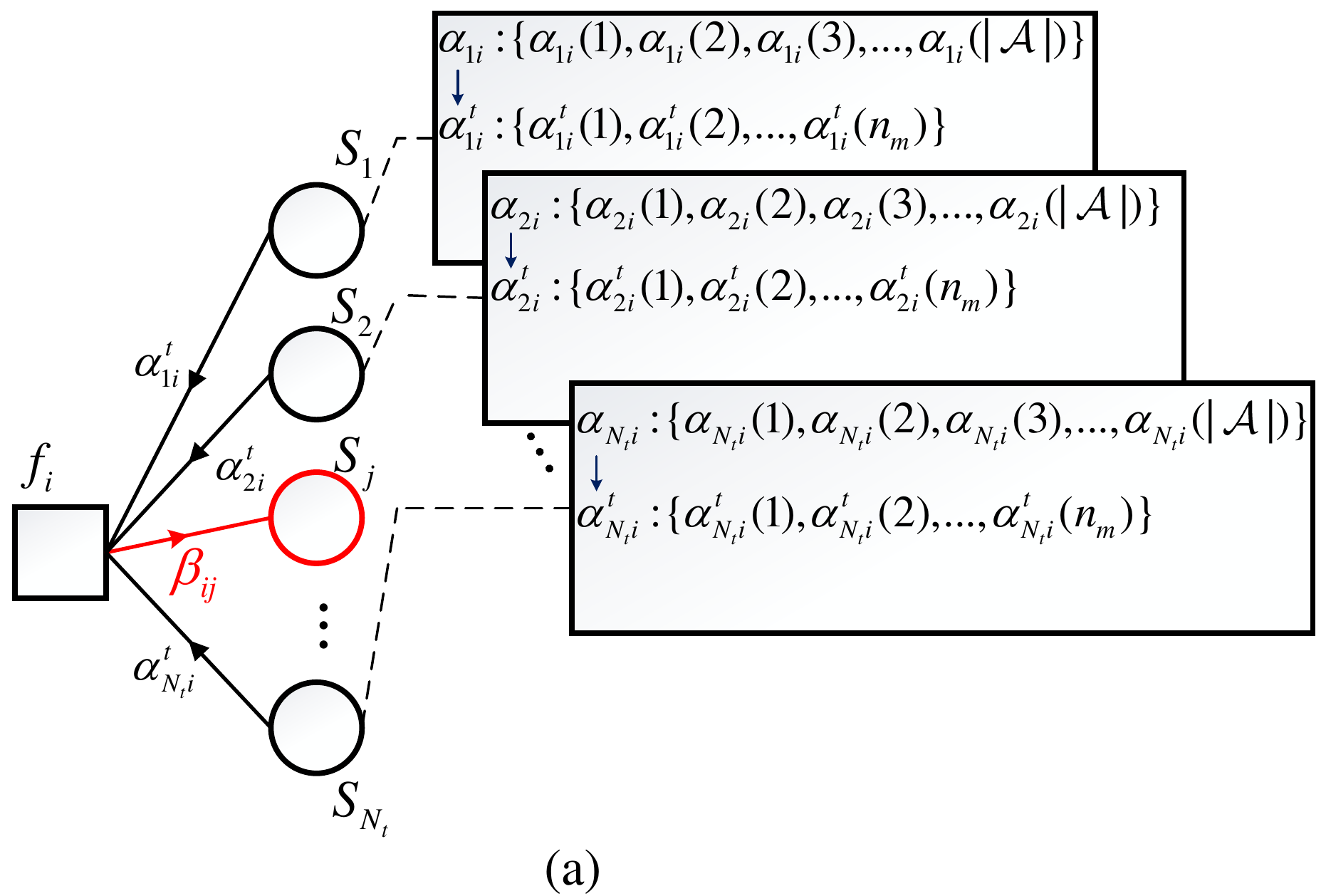}
    \label{fig:configSCR}
    \end{minipage}
  }
  \subfigure{
    \begin{minipage}[t]{0.4\linewidth}
    \centering
    \includegraphics[width=\linewidth]{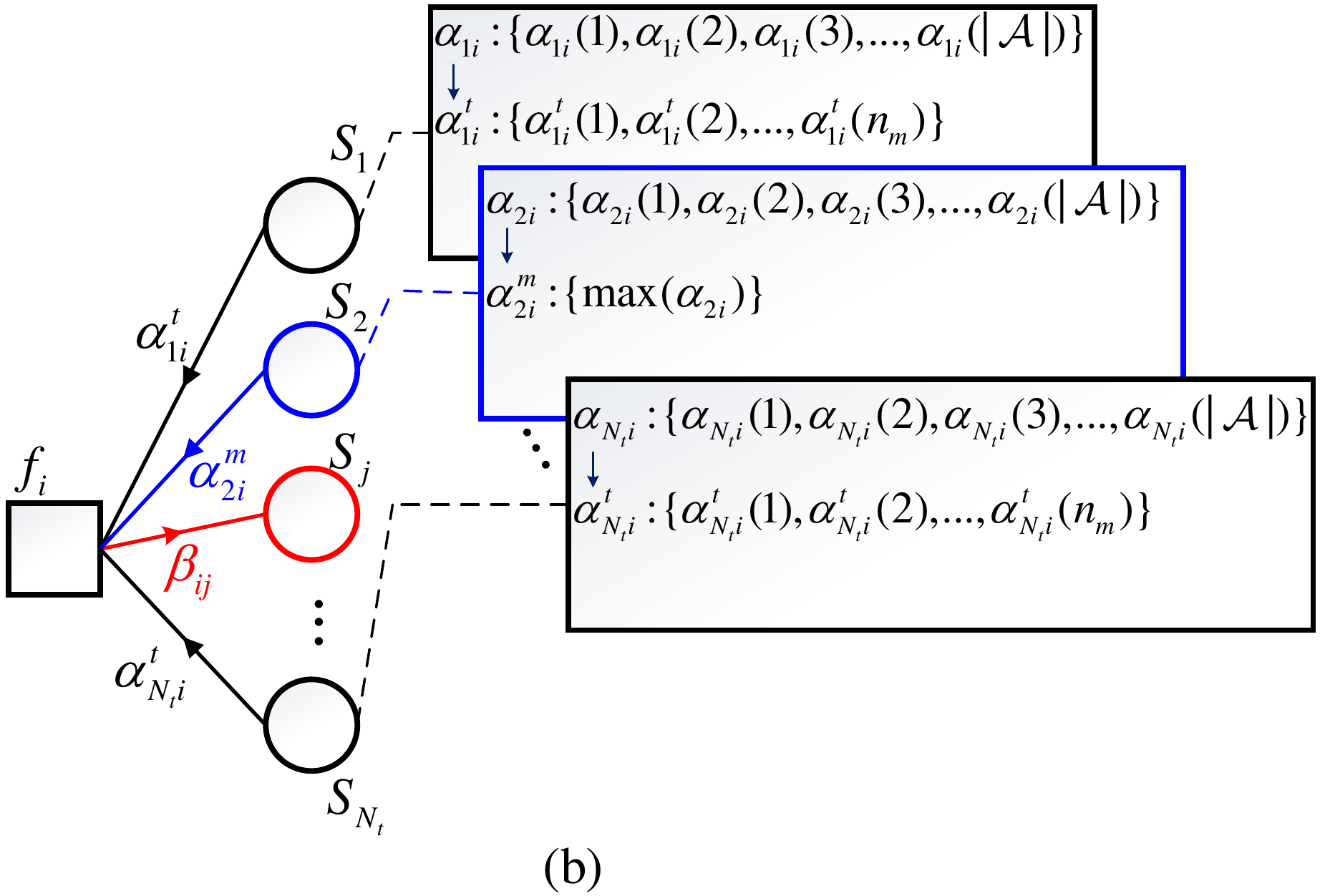}
    \label{fig:configECR}
    \end{minipage}
  }
  \caption{(a): the message update of $\beta_{ij}$ with the ST strategy; \quad (b) the message update of $\beta_{ij}$ with both the ST strategy and the ES strategy.}
\end{figure*}

    \subsection{Symbol-Based Truncation Strategy}\label{sec:STS}
The basic idea of the ST strategy is that each received $\alpha$ message can be truncated by eliminating the components with relatively small values. Then the $\beta$ messages are updated with the truncated $\alpha$ messages, contributing to a squeezed search space of the transmitted symbol vector.
Fig.~\ref{fig:configSCR} shows the details of the $\beta_{ij}$ message update with the ST strategy.
It is observed that each symbol node $S_k$ ($k\neq j$) associates with a rectangle box involving the incoming message $\alpha_{ki}$. Each incoming message is comprised of $|\mathcal{A}|$ symbol LLRs. Then, the symbol LLRs in message $\alpha_{ki}$ are sorted to build up the truncated messages $\alpha^t_{ki}$ with the largest $d_m$ ($d_m\ll|\mathcal{A}|$) symbol LLRs. Let $\mathbf{s}_{\mathbf{{k}}/j}$ represent the subvector of transmitted symbol vector $\mathbf{s}$ except the transmitted symbol $s_j$. Denote $\mathcal{B}(d_m)$ as a configuration set consisting of all of the possible choices of the $\mathbf{s}_{\mathbf{{k}}/j}$. With the truncated messages, $\mathcal{B}(d_m)$ is illustrated as
\begin{equation}\label{eq:configSet}
\begin{aligned}
\mathcal{B}(d_m) &= \Bigg \{ \mathbf{s}_{\mathbf{k}/j} = \Big[\mu_{k_1}^1, \mu_{k_2}^2, ..., \mu_{k_{N_t-1}}^{N_t-1}\Big]^T: \\
& \forall\mathbf{k}=[k_1, k_2, ...,k_{N_t-1}]^T \in \{1,2,...,d_m\}^{N_t-1} \Bigg \},
\end{aligned}
\end{equation}
where $\mu_{k_l}^l$ ($l\in\{1,2,...,N_t-1\}$) is the $k_l$-th transmitted symbol in the constellation dominated by the truncated messages $\alpha^t_{k_l i}$.
With the configuration set $\mathcal{B}(d_m)$, the formula for the message update of $\beta_{ij}$ is represented as (\ref{eq:betaMessageEMS}).
\begin{figure*}[ht]
\normalsize
\setcounter{equation}{8}
\begin{equation}\label{eq:betaMessageEMS}
\begin{aligned}
\beta_{ij}^{(l)}(k)  =
\mathop{{\mathrm{max}}}_{\mathbf{s}: s_j=\mu_k, \atop \mathbf{s}_{\mathbf{k}/j} \in \mathcal{B}(d_m)} \left \{\! -\frac{1}{2\sigma^2}\|y_i - \mathbf{h}_i \mathbf{s}\|^2 + \! \mathop{\sum}_{t=1,t\neq j}^{N_t} \alpha_{ti}^{(l-1)} \right \}
 - \mathop{\mathrm{max}}_{\mathbf{s}: s_j=\mu_1, \atop \mathbf{s}_{\mathbf{k}/j} \in \mathcal{B}(d_m)}
\left \{\! -\frac{1}{2\sigma^2}\|y_i - \mathbf{h}_i \mathbf{s}\|^2 + \! \mathop{\sum}_{t=1,t\neq j}^{N_t} \alpha_{ti}^{(l-1)} \right \}.
\end{aligned}
\end{equation}
\setcounter{equation}{9}

\hrulefill
\vspace*{4pt}

\end{figure*}

It is emphasized that the cardinality of the configuration set $\mathcal{B}(d_m)$ is $|\mathcal{B}(d_m)|=(d_m)^{(N_t-1)}$.  With this cardinality, the search space of the transmitted symbol vector in (\ref{eq:betaMessageEMS}) has a size of $|\mathcal{A}|\times|\mathcal{B}(d_m)|$.
This search space is smaller than the search space in (\ref{eq:simpleBeta}). Therefore, the BsP detection with the ST strategy enjoys a lower computational complexity compared with the Original-BP detection.

       \subsection{Edge-Based Simplification Strategy}\label{sec:ESS}
Although the ST strategy brings down the base of the exponential associated with the search space of the transmitted symbol vector, the BsP detection still suffers from an exponentially increasing computational complexity dominated by the number of the transmit antennas. In order to further reduce the computational complexity of the BsP detection, the ES strategy is proposed in this subsection. With the ES strategy, some of the transmitted symbols in $\mathbf{s}_{\mathbf{k}/j}$ are directly decided as the symbols with the largest reliability according to its related $\alpha$ messages.
In the message update of $\beta_{ij}$, given $N_t-1$ incoming edges, it is assume that there are $d_f-1$ \textit{chosen edges}, in which the associated transmitted symbols are decided with the truncated massages $\alpha^t_{ki}$. Consequently, there are the other $N_t-d_f$ \textit{simplified edges}. The transmitted symbols linked with these simplified edges are directly decided as the symbols with the largest reliability. The incoming $\alpha$ messages only involving the largest symbol LLR are denoted as message $\alpha^m_{ki}$.
Let $\mathcal{B}(d_m, d_f)$ represent the subset of the configuration set $\mathcal{B}(d_m)$ with $d_f-1$ chosen edges, then $\mathcal{B}(d_m, d_f)$ is illustrated as
\begin{equation}\label{eq:configSubset}
\mathcal{B}(d_m, d_f) = \left \{ \mathbf{s}_{\mathbf{k}/j}=\left[ \mathbf{s}^T_{\mathbf{k}/j,d_f},\ \mathbf{\bar{s}}^T_{\mathbf{k}/j,d_f} \right]^T \right\},
\end{equation}
where $\mathbf{s}_{\mathbf{k}/j,d_f}$ is comprised of the transmitted symbols corresponding to the $d_f-1$ chosen edges. Therefore the expression of $\mathbf{s}_{\mathbf{k}/j,d_f}$ is
\begin{equation}\label{eq:configSubsetDf}
\begin{aligned}
\mathbf{s}_{\mathbf{k}/j,d_f}  =  &\left[\mu_{k1}^1, \mu_{k2}^2,...,\mu_{k_{d_f-1}}^{d_f-1} \right]^T: \\
& \ \ \ \ \ \ \ \; \forall\mathbf{k}=\left\{1,2,...,d_m \right\}^{d_f-1}.
\end{aligned}
\end{equation}
The notation $\mathbf{\bar{s}}_{\mathbf{k}/j,d_f}$ in (\ref{eq:configSubset}) represents the complementary transmitted symbol set of the $\mathbf{s}_{\mathbf{k}/j,d_f}$ in terms of the $\mathbf{s}_{\mathbf{k}/j}$.
Each component in $\mathbf{\bar{s}}_{\mathbf{k}/j,d_f}$ is the transmitted symbol corresponding to the largest symbol LLR preserved in the message $\alpha^m_{ki}$.
Note that the chosen edges are the arbitrary $d_f-1$ edges selected from the $N_t-1$ incoming edges, therefore the cardinality of the configuration set $\mathcal{B}(d_m,d_f)$ is $|\mathcal{B}(d_m,d_f)|=\binom{N_t-1}{d_f-1}d_m^{(d_f-1)}$.
Compared with the configuration set $\mathcal{B}(d_m)$, the $\mathcal{B}(d_m,d_f)$ enjoys a smaller cardinality by bringing down the exponent of the exponential for the search space size from $N_t-1$ to $d_f-1$.
The Original-BP detection requires to search $|\mathcal{A}|^{N_t}$ possible transmitted symbol vectors to update the message $\beta_{ij}$. Compared with the Original-BP detection, the BsP detection with the configuration set $\mathcal{B}(d_m,d_f)$ only needs to search $|\mathcal{A}|\binom{N_t-1}{d_f-1}d_m^{(d_f-1)}$ possible transmitted symbol vectors to update the message $\beta_{ij}$. Therefore, compared with the Original-BP detection, the BsP detection enjoys a lower computational complexity. The computational complexity of the BsP detection with the ST strategy and the ES strategy will be further discussed in Section~\ref{sec:Complexity}.

Fig.~\ref{fig:configECR} shows the message update of $\beta_{ij}$ with the ST strategy and the ES strategy in the BsP detection.
The blue symbol nodes corresponding to the simplified edges are linked with the blue boxes involving the message $\alpha^m_{ki}$.
The black symbol nodes corresponding to the chosen edges are connected to the black boxes containing the message $\alpha^t_{ki}$. With the configuration set $\mathcal{B}(d_m,d_f)$, the $\beta$ message in the BsP detection is updated as (\ref{eq:betaMessageEMSDf}).
\begin{figure*}[t]
\normalsize
\setcounter{equation}{11}
\begin{equation}\label{eq:betaMessageEMSDf}
\begin{aligned}
\beta_{ij}^{(l)}(k)  =
\mathop{{\mathrm{max}}}_{\mathbf{s}: s_j=\mu_k, \atop \mathbf{s}_{\mathbf{k}/j} \in \mathcal{B}(d_m,d_f)} \left \{\! -\frac{1}{2\sigma^2}\|y_i - \mathbf{h}_i \mathbf{s}\|^2 + \! \mathop{\sum}_{t=1,t\neq j}^{N_t} \alpha_{ti}^{(l-1)} \right \}
 - \mathop{\mathrm{max}}_{\mathbf{s}: s_j=\mu_1, \atop \mathbf{s}_{\mathbf{k}/j} \in \mathcal{B}(d_m,d_f)}
\left \{\! -\frac{1}{2\sigma^2}\|y_i - \mathbf{h}_i \mathbf{s}\|^2 + \! \mathop{\sum}_{t=1,t\neq j}^{N_t} \alpha_{ti}^{(l-1)} \right \}.
\end{aligned}
\end{equation}

\setcounter{equation}{12}
\hrulefill
\vspace*{4pt}

\end{figure*}

    \subsection{Initialization with PP Information}\label{sec:init}
The BsP detection requires the \textit{a priori} information for each transmitted symbol to build up the configuration set.
However, this \textit{a priori} information is unavailable (set as the zero information) when the BsP detection starts to update the $\beta$ messages. A natural way to obtain the initial \textit{a priori} information is to perform one iteration of the Original-BP detection, but this way incurs an unacceptable computational complexity, which is contrary to the motivation of the proposed BsP detection.

Recently, many works \cite{goldberger2010pseudo, Feichi11_SS} utilize the PP information generated by a simple linear detector to initialize the \textit{a priori} information, making the BP detection converge fast. In \cite{Feichi11_SS}, an LMMSE detector is employed to initialize the \textit{a priori} information for the GF-BP detection, resulting in an enhanced error rate performance.
Inspired by this, the LMMSE detector is utilized to provide the initial \textit{a priori} information for the proposed BsP detection. The details are as follows.

The LMMSE detector estimates the transmitted symbol vector by
\begin{equation}\label{eq:MMSE}
\mathbf{\hat{s}}_{\mathrm{MMSE}} = (\mathbf{H}^H\mathbf{H}+\sigma^2\mathbf{I})^{-1}\mathbf{H}^H\mathbf{y}.
\end{equation}
With the results of the LMMSE detector, the \textit{a priori} probabilities of transmitted symbols are computed by
\begin{equation}\label{eq:priorProb}
p_j(\mu_k) = \mathrm{exp}\left\{ -\frac{\|\mu_k-\mathbf{\hat{s}}_{\mathrm{MMSE}_j}\|^2}{2\sigma^2_{\mathrm{MMSE}_j}} \right\},
\end{equation}
where $\mathbf{\hat{s}}_{\mathrm{MMSE}_j}$ is the estimates of the $j$-th ($j\in \{1,2,...,N_t\}$) symbol, and $\sigma^2_{\mathrm{MMSE}_j}$ is the element of the $j$-th row and $j$-th column in the noise covariance matrix $\mathbf{K}$. The matrix $\mathbf{K}$ is calculated by
\begin{equation}\label{eq:covarianceMtx}
\mathbf{K} = (\mathbf{H}^H\mathbf{H}+\sigma^2\mathbf{I})^{-1}.
\end{equation}
Given the \textit{a priori} probability for each transmitted symbol, the initial $\alpha$ messages are computed according to (\ref{eq:LLR}).

    \subsection{The BsP Detection Algorithm}\label{sec:alg}
\IncMargin{1em}
\begin{algorithm}[htpb]
\SetKwInOut{Input}{Input}\SetKwInOut{Output}{Output}\SetKwFunction{MMSE}{InitMMSE}

    \Input{received signal $\mathbf{y}$, channel matrix $\mathbf{H}$}
    \Output{soft information of coded bits $\mathbf{r}$}
    \BlankLine

    \emph{Initialization}: $\bm{\alpha} \leftarrow$ \MMSE{$\mathbf{y}$, $\mathbf{H}$}\;
    \For{ $Iter \leftarrow 1$ \KwTo $\mathrm{MaxIterNum}$}{

        \For(\tcp*[h]{compute $\bm{\beta}$ messages}){ $i \leftarrow 1$ \KwTo $N_r$}{

            \For(\tcp*[h]{compute truncated $\alpha$ messages}){$j \leftarrow 1$ \KwTo $N_t$}{
                \texttt{sort symbol LLRs in} $\alpha_{ji}$ \;
                \texttt{reserve the largest $d_m$ symbol LLRs to form $\alpha_{ji}^t$} \;
            }

            \For{$j \leftarrow 1$ \KwTo $N_t$}{
                \texttt{build} $\mathcal{B}(d_m,d_f)$ \texttt{as} (\ref{eq:configSubset})\;
                \texttt{update} $\beta_{ij}$ \texttt{as} (\ref{eq:betaMessageEMSDf}) \;
            }

        }
        \For(\tcp*[h]{compute $\bm{\alpha}$ messages}){$j \leftarrow 1$ \KwTo $N_t$}{
            \texttt{update} $\gamma_j$ \texttt{as} (\ref{eq:output}) \;
            \For{$i \leftarrow 1$ \KwTo $N_r$}{
                $\alpha_{ji} \leftarrow \gamma_j - \beta_{ij}$ \;
            }
        }
    }
    \texttt{compute} $\mathbf{r}$ \texttt{as} (\ref{eq:newBitLLR}) \;

\caption{BsP Detection}
\label{Alg:EMS_BP}
\end{algorithm}
\DecMargin{1em}

Algorithm~\ref{Alg:EMS_BP} shows the BsP detection algorithm with the configuration set $\mathcal{B}(d_m,d_f)$. The \textit{a priori} information is initialized with an LMMSE detector. The details are as follows.
Given the received signal $\mathbf{y}$ and the channel matrix $\mathbf{H}$, the LMMSE detector is utilized first to initialize the \textit{a priori} probabilities of the transmitted symbols as (\ref{eq:priorProb}).
Then the initial $\alpha$ messages are acquired as (\ref{eq:LLR}).
In order to update the $\beta$ messages, for each factor node $f_i$, the symbol LLRs in each incoming message $\alpha_{ji}$ ($j\in \{1,2,...,N_t\}$) are sorted to build up the truncated message $\alpha^t_{ji}$ with the  largest $d_m$ symbol LLRs.
Note that the message $\alpha^t_{ji}$ in the message update of $\beta_{ij}$ ($j\in \{1,2,...,N_t\}$) for the factor node $f_i$ can be reused, therefore the symbol LLRs of each incoming $\alpha$ message only needs to be sorted once in one iteration. With the truncated message $\alpha^t_{ji}$, the configuration set $\mathcal{B}(d_m,d_f)$ is built up as (\ref{eq:configSubset}). The message $\beta_{ij}$ is then updated as (\ref{eq:betaMessageEMSDf}).
To update the $\alpha$ messages, the incoming  messages $\beta_{ij}$ ($i\in \{1,2,...,N_r\}$) for each symbol node $S_j$ are added together to acquire the overall symbol LLRs. Then the message $\alpha_{ji}$ is updated with
\begin{equation}\label{eq:alphaUpdate}
\alpha_{ji}(k) = \gamma_j(k) - \beta_{ij}(k).
\end{equation}
The BsP detection stops the message update when the maximum iteration number is achieved. Finally, it outputs the bit LLRs computed as (\ref{eq:newBitLLR}).

{The proposed BsP detection achieves a lower computational complexity by shrinking the search space of the transmitted symbol vector with the ST strategy and the ES strategy.
Empirically, seeing that the shrunk search space may miss the optimal transmitted symbol vector, the error rate performance of the BsP detection will be penalized. However, experiments in Section~\ref{sec:Results} demonstrate different results. The reason is comprised of two parts. On the one hand, the BsP detection with the shrunk search space circumvents the propagation of the messages with relatively small probabilities, weakening the influence of the loops in the FG. On the other hand, the BsP detection can benefit from the more precise initial \textit{a priori} information provided by an LMMSE detector for each transmitted symbol. Such an effective initialization brings down the probability that the shrunk search space loses the the optimal transmitted symbol vector.}
	
    \section{Computational Complexity}\label{sec:Complexity}
In this section, the computational complexity of the proposed BsP detection is analyzed. Then this complexity is compared with the the complexity of the SOA BP detection algorithms, involving the the Original-BP detection and the EBRDF-BP detection in \cite{Jun08_JSAC}, the PP-GF-BP detection \cite{Feichi11_SS}, and the RD-GAI-BP detection in \cite{yang2018low}.

    \subsection{Complexity Analysis of the BsP Detection}
For fair comparison, the computational complexity for each detection algorithm is measured with the principle in \cite{Jun08_JSAC}. This principle instructs that the computational complexity of the BP detection is dominated by the number of multiplication operations per channel use (deliver $MN_t$ transmission bits), because the multiplication operation demands much higher computational complexity than the addition and comparison operations.
Therefore, the computational complexity of the proposed BsP detection is represented with the number of the multiplication operation in this section.
For clarity, the computational complexity is analyzed with the following rules:
\begin{enumerate}
  \item The multiplication operation results corresponding to the $\mathbf{Hs}$ are reused in different iterations.
  \item One complex number multiplication operation requires four real number multiplication operations.
\end{enumerate}

Since the proposed BsP detection with the configuration set $\mathcal{B}(|\mathcal{A}|,N_t)$ shares the same computational complexity with the Original-BP detection when updating the $\beta$ messages in each iteration, its computational complexity can be derived from the computational complexity of the Original-BP detection. According to \cite{Jun08_JSAC}, the Original-BP detection requires to search $|\mathcal{A}|^{N_t}$ possible transmitted symbol vector $\mathbf{s}$ to compute the $\mathbf{h}_i\mathbf{s}$ in (\ref{eq:simpleBeta}), thus suffering from a computational complexity of $4|\mathcal{A}|^{N_t}N_rN_t$.
Note that this computational complexity is irrelevant to the iteration number $Q_L$, because the computational results of the $\mathbf{h}_i\mathbf{s}$ are reused in different iterations.
In the same way, the BsP detection with configuration set $\mathcal{B}(d_m,d_f)$ requires to search $|\mathcal{A}|\binom{N_t-1}{d_f-1}d_m^{(d_f-1)}$  possible transmitted symbol vector $\mathbf{s}$ to compute the $\mathbf{h}_i\mathbf{s}$ in (\ref{eq:output}). Therefore, the overall computational complexity of the BsP with $\mathcal{B}(d_m,d_f)$ is $4|\mathcal{A}|\binom{N_t-1}{d_f-1}d_m^{(d_f-1)}N_tN_r$ (for simplicity, the additional complexity of the initialization is ignored here).
Note that, although the BsP detection requires the additional $d_m|\mathcal{A}|N_rN_tQ_L$ comparison operations to sort the symbol LLRs for each incoming $\alpha$ message, the computational complexity of the comparison operations is still insignificant compared with the complexity of the multiplication operations.

    \subsection{Complexity Comparisons with the SOA BP Detections}

\renewcommand{\arraystretch}{1.5} 
\begin{table*}[t]
	\caption{Complexity Comparisons of the Proposed BsP and the SOA BP Detection Algorithms}
	\centering
	\scalebox{0.8}{
        \begin{threeparttable}[b]
		\begin{tabular}{ l | c | c | c | c }
			\Xhline{1.2pt}
			\diagbox{\textbf{Solutions}}{\textbf{Operations}} &\textbf{Multiplications} &\textbf{Additions} &\textbf{Comparisons} &\textbf{Exponentiations} \\
            \hline
            \hline
            \textbf{Original-BP} \cite{Jun08_JSAC}  &{$\mathcal{O}( |\mathcal{A}|^{N_t}N_tN_r ) $} &{\makecell[c]{$\mathcal{O}(|\mathcal{A}|^{N_t}N_tN_r)+$ \\ $\mathcal{O}(|\mathcal{A}|^{N_t-1}N_rN_t^2Q_L)$ }} &{$\mathcal{O}(|\mathcal{A}|^{N_t}N_tN_rQ_L) $} &$0$   \\
            \hline
            \textbf{EBRDF-BP} \cite{Jun08_JSAC} &{$\mathcal{O}(|\mathcal{A}|^{d_f}d_fN_r)$} &{\makecell[c]{$\mathcal{O}(|\mathcal{A}|^{d_f}d_fN_r)$ \\ $\mathcal{O}(|\mathcal{A}|^{d_f-1}N_rN_td_fQ_L)$}} &{$\mathcal{O}(|\mathcal{A}|^{d_f}N_tN_rQ_L)$} &$0$   \\
            \hline
            \textbf{PP-GF-BP$^{\bm{\dag}}$} \cite{Feichi11_SS} &{$\mathcal{O}(|\mathcal{A}|N_tN_rQ_L)+\mathcal{O}(N_t^3)$} &{$\mathcal{O}(|\mathcal{A}|N_tN_rQ_L)+\mathcal{O}(N_t^3)$} &{$0$} &{$\mathcal{O}(|\mathcal{A}|N_tN_rQ_L)$}\\
            \hline
            \textbf{RD-GAI-BP} \cite{yang2018low} &{$\mathcal{O}(\sqrt{|\mathcal{A}|}N_tN_rQ_L)$} &{$\mathcal{O}(\sqrt{|\mathcal{A}|}N_tN_rQ_L)$} &{$0$} &{$\mathcal{O}(\sqrt{|\mathcal{A}|}N_tN_rQ_L)$} \\
            \hline
            \textbf{Proposed BsP$^{\bm{\dag}}$} &{$\mathcal{O}(|\mathcal{A}|\binom{N_t-1}{d_f-1}d_m^{(d_f-1)}N_tN_r)+\mathcal{O}(N_t^3)$}
            &{\makecell[c]{$\mathcal{O}(|\mathcal{A}|\binom{N_t-1}{d_f-1}d_m^{(d_f-1)}N_tN_r)+$ \\ $\mathcal{O}(\binom{N_t-1}{d_f-1}d_m^{(d_f-1)}N_tN_rQ_L)+\mathcal{O}(N_t^3)$}}
            &{\makecell[c]{ $\mathcal{O}(d_m|\mathcal{A}|N_rN_tQ_L)+$\\
            $\mathcal{O}(|\mathcal{A}|\binom{N_t-1}{d_f-1}d_m^{(d_f-1)}N_tN_rQ_L)$}}
            &{$0$} \\
			\Xhline{1.2pt}

	\end{tabular}
    \begin{tablenotes}
    \item[${\bm{\dag}}$]: Initialized with an LMMSE detector.
    \end{tablenotes}
    \end{threeparttable}

}
	\label{table:ComplexityComparison}
\end{table*}

Table~\ref{table:ComplexityComparison} shows the computational complexity of the proposed BsP detection and the state-of-the art BP detection algorithms. For simplicity, big $\mathcal{O}$ representation is used in Table~\ref{table:ComplexityComparison}. Several observations can be seen from the table.
First, the Original-BP detection suffers from the highest computational complexity.
This computational complexity is even comparable to that of the MAP detection.
The computational complexity of the EBRDF-BP detection is dominated by the $d_f$.
It is noted that although the EBRDF-BP detection can achieve a lower computational complexity with a smaller $d_f$, it sacrifices too much error rate performance.
Second, the PP-GF-BP detection and the RD-GAI-BP detection circumvent the exponentially increasing complexity, despite its requirement of exponentiation operations.
Finally, the proposed BsP detection has the computational complexity of $\mathcal{O}(|\mathcal{A}|\binom{N_t-1}{d_f-1}d_m^{(d_f-1)}N_tN_r)+\mathcal{O}(N_t^3)$, where $\mathcal{O}(N_t^3)$ is the additional complexity of the LMMSE detector.
This computational complexity is dominated by the $d_m$ and the $d_f$.
Empirically, the BsP detection with $\mathcal{B}(2,2)$ exhibits a good enough error rate performance, which is demonstrated in Section~\ref{subsec:differentSets}. Therefore, it also circumvents the exponentially increasing complexity in practical MIMO systems.
Compared with the Original-BP detection and the EBRDF-BP detection, the BsP achieves a lower computational complexity. Compared with the PP-GF-BP detection and the RD-GAI-BP detection, the
BsP detection reaches a comparable computational complexity.

\begin{figure}[t]
  \centering
  \includegraphics[width=0.9\linewidth]{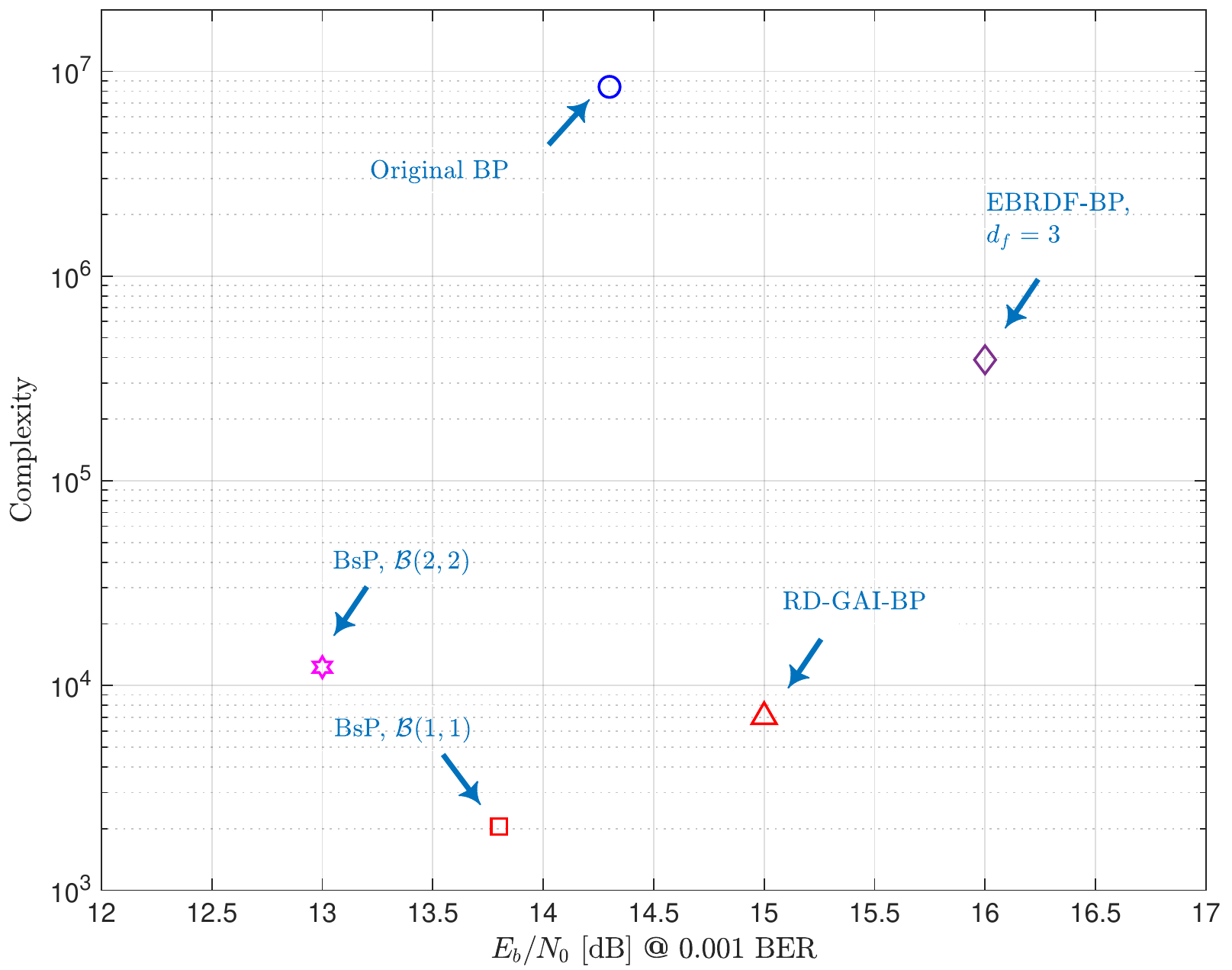}
  \caption{Complexity vs. error rate performance for the proposed BsP detection and the SOA BP detection algorithms in the MIMO system with $N_r=8$, $N_t=4$ and $16$-QAM modulation.}
  \label{fig:complexity}
\end{figure}
Fig.~\ref{fig:complexity} shows the comparison results of the computational complexity vs. error rate performance for the proposed BsP detection and the SOA BP detection algorithms in the MIMO system with $N_r=8$, $N_t=4$, and $16$-QAM modulation.
It is observed that the Original-BP detection suffers from the highest computational complexity while the EBRDF-BP detection suffers from the worst BER performance.
The BsP detection with the configuration set $\mathcal{B}(2,2)$ shows the best BER performance, although its computational complexity is slightly higher than that of the RD-GAI-BP detection. The BsP detection with the configuration set $\mathcal{B}(1,1)$ enjoys the lowest computational complexity, and achieves a better error rate performance than the RD-GAI-BP detection.
Here note that the configuration set $\mathcal{B}(1,1)$ indicates that there is no chosen edge when updating the message $\beta_{ij}$, and the $\mathbf{s}_{\mathbf{k}/j}$ in (\ref{eq:betaMessageEMSDf}) is comprised of the transmitted symbols associated with the largest symbol LLR in each incoming message $\alpha_{ji}$.
In fact, if either of the $d_m$ and $d_f$ is equal to one, the configuration set $\mathcal{B}(d_m,d_f)$ involves the only $\mathbf{s}_{\mathbf{k}/j}$ consisting of the transmitted symbols dominated by each $\alpha^m_{ji}$ message.
In addition, the PP-GF-BP detection is absent from the figure, due to its poor error rate performance.
Fig.~\ref{fig:complexity} shows that the BsP detection realizes the trade-off between the computational complexity and the error rate performance by configuring different ST and ES parameters. Further more, compared with the SOA BP detection algorithms, it succeeds in reducing the computational complexity and achieving a better error rate performance.

	\section{Numerical Results}\label{sec:Results}
In this section, the numerical results are provided to demonstrate the error rate performance of the BsP detection. It is assumed that the channel matrix are perfectly recovered in the receiver. The error rate performance of the BsP detection and the SOA BP detection algorithms are compared in the scenarios as:
1) the MIMO system with $N_r=8$, $N_t=4$ and $16$-QAM modulation, 2) the MIMO system with $N_r=8$, $N_t=4$ and $64$-QAM modulation, and 3) the MIMO system with $N_r=16$, $N_t=8$ and $16$-QAM modulation. For fair comparison, the number of the iterations is set as $Q_L=10$ for all of the BP detection.

    \subsection{Comparison Results with the SOA BP Detections}
In this subsection, the proposed BsP detection is compared with the the Original-BP detection and the EBRDF-BP detection in \cite{Jun08_JSAC}, and the RD-GAI-BP detection in \cite{yang2018low}.
The GAI-BP detection in \cite{som2011low} and the PP-GF-BP detection in  \cite{Feichi11_SS} are not considered due to its aggravated performance error floor in MIMO systems with high-order modulations.

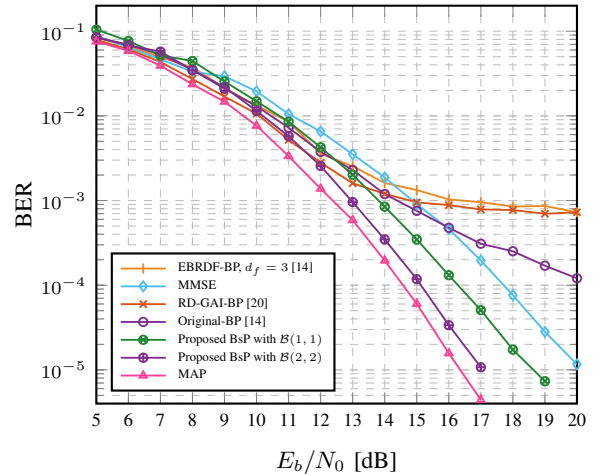
\begin{figure}[htbp]
  \centering
  \begin{tikzpicture}
    \definecolor{myblued}{RGB}{0,114,189}
    \definecolor{myred}{RGB}{217,83,25}
    \definecolor{myyellow}{RGB}{237,137,32}
    \definecolor{mypurple}{RGB}{126,47,142}
    \definecolor{myblues}{RGB}{77,190,238}
    \definecolor{mygreen}{RGB}{32,134,48}
    \definecolor{mypink}{RGB}{255,62,150}
      \pgfplotsset{
        label style = {font=\fontsize{9pt}{7.2}\selectfont},
        tick label style = {font=\fontsize{7pt}{7.2}\selectfont}
      }

    \begin{axis}[
        scale = 1,
        ymode=log,
        xmin=5.0,xmax=20,
        ymin=4.0E-06,ymax=0.2,
        xlabel={$E_b/N_0$ [dB]}, xlabel style={yshift=0.1em},
        ylabel={BER}, ylabel style={yshift=-0.75em},
        xtick={5,6,7,8,9,10,11,12,13,14,15,16,17,18,19,20},
        xticklabels={5,6,7,8,9,10,11,12,13,14,15,16,17,18,19,20},
        grid=both,
        ymajorgrids=true,
        xmajorgrids=true,
        grid style=dashed,
        width=0.9\linewidth,
        thick,
        mark size=1,
        legend style={
          nodes={scale=0.75, transform shape},
          anchor={center},
          cells={anchor=west},
          column sep= 1mm,
          row sep= -0.25mm,
          font=\fontsize{6.5pt}{7.2}\selectfont,
        },
        legend columns=1,
        legend pos=south west,
    ]

    \addplot[
        color=myyellow,
        mark=|,
        line width=0.25mm,
        mark size=1.9,
        fill opacity=0,
    ]
    table {
      5  0.0833333
      6  0.0668305
      7  0.0491071
      8  0.0375564
      9  0.021271
      10 0.0132584
      11 0.00858362
      12 0.00355764
      13 0.00253954
      14 0.00161183
      15 0.001329
      16 0.00103058
      17 0.00095598
      18 0.000850203
      19 0.000861482
      20 0.000726779
    };
    \addlegendentry{EBRDF-BP, $d_f=3$ \cite{Jun08_JSAC}}

    \addplot[
        color=myblues,
        mark=diamond*,
        line width=0.25mm,
        mark size=1.9,
        fill opacity=0,
    ]
    table {
      5  0.0887097
      6  0.0639563
      7  0.0465468
      8  0.0333126
      9  0.0293605
      10 0.0194706
      11 0.0104891
      12 0.00655492
      13 0.00351919
      14 0.00187229
      15 0.000902961
      16 0.000461424
      17 0.000195836
      18 7.62338e-05
      19 2.81196e-05
      20 1.16196e-05
    };
    \addlegendentry{MMSE}

    \addplot[
        color=myred,
        mark=x,
       line width=0.25mm,
        mark size=1.9,
        fill opacity=0,
    ]
    table {
      5  0.0789
      6  0.0612
      7  0.0432
      8  0.0272
      9  0.0169
      10 0.0106
      11 0.0052
      12 0.0028
      13 0.0016
      14 0.0012
      15 9.5170e-04
      16 8.8251e-04
      17 7.8499e-04
      18 7.7223e-04
      19 6.9740e-04
      20 7.2688e-04
    };
    \addlegendentry{RD-GAI-BP \cite{yang2018low}}

    \addplot[
        color=mypurple,
        mark=*,
        line width=0.25mm,
        mark size=1.6,
        fill opacity=0,
    ]
    table {
      5   0.0844957
      6   0.0691038
      7   0.0528486
      8   0.0350131
      9   0.0209269
      10  0.0130154
      11  0.00724302
      12  0.00376242
      13  0.0022977
      14  0.00119105
      15  0.000755619
      16  0.000476702
      17  0.000308966
      18  0.000251631
      19  0.000169889
      20  0.000120675
    };
    \addlegendentry{Original-BP \cite{Jun08_JSAC}}

     \addplot[
        color=mygreen,
        mark=otimes*,
        line width=0.25mm,
        mark size=1.6,
        fill opacity=0,
    ]
    table {
      5  0.104432
      6  0.0765957
      7  0.0518771
      8  0.0444236
      9  0.0254859
      10 0.0148458
      11 0.00853277
      12 0.00425407
      13 0.00201504
      14 0.000845318
      15 0.000346684
      16 0.000131103
      17 5.0784e-05
      18 1.73276e-05
      19 7.34441e-06

    };
    \addlegendentry{Proposed BsP with $\mathcal{B}(1,1)$}

    \addplot[
        color=mypurple,
        mark=oplus*,
        line width=0.25mm,
        mark size=1.6,
        fill opacity=0,
    ]
    table {
      5   0.0844933
      6   0.0681497
      7   0.0573657
      8   0.0346091
      9   0.0218156
      10  0.0114301
      11  0.00574092
      12  0.00256042
      13  0.00095799
      14  0.000347315
      15  0.000117874
      16  3.36968e-05
      17  1.07373e-05
    };
    \addlegendentry{Proposed BsP with $\mathcal{B}(2,2)$}

    \addplot[
        color=mypink,
        mark=triangle*,
       line width=0.25mm,
        mark size=1.6,
        fill opacity=0,
    ]
    table {
      5  0.0755965
      6  0.0585664
      7  0.0391406
      8  0.02358
      9  0.0146888
      10 0.00764489
      11 0.00334853
      12 0.00137913
      13 0.000586802
      14 0.000195404
      15 6.03787e-05
      16 1.57112e-05
      17 4.4188e-06
    };
    \addlegendentry{MAP}

    \end{axis}
    \end{tikzpicture} \\
  \caption{BER Performance of the BsP detection and the SOA BP detection algorithms in the MIMO system with $N_r=8$, $N_t=4$ and $16$-QAM modulation.}
  \label{fig:performanceTx4Rx8}
\end{figure}

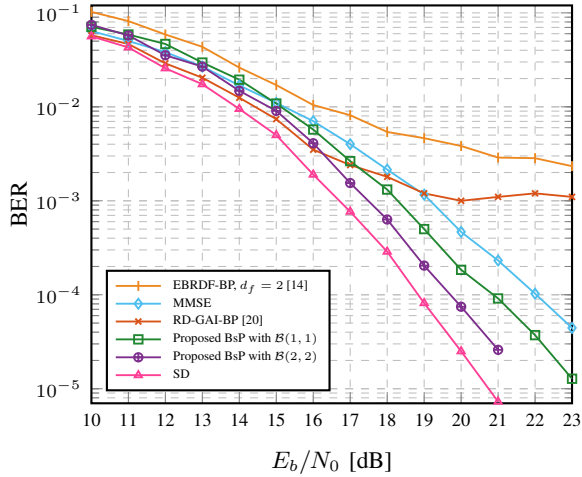
\begin{figure}[htbp]
  \centering
  \begin{tikzpicture}
    \definecolor{myblued}{RGB}{0,114,189}
    \definecolor{myred}{RGB}{217,83,25}
    \definecolor{myyellow}{RGB}{237,137,32}
    \definecolor{mypurple}{RGB}{126,47,142}
    \definecolor{myblues}{RGB}{77,190,238}
    \definecolor{mygreen}{RGB}{32,134,48}
    \definecolor{mypink}{RGB}{255,62,150}
      \pgfplotsset{
        label style = {font=\fontsize{9pt}{7.2}\selectfont},
        tick label style = {font=\fontsize{7pt}{7.2}\selectfont}
      }

    \begin{axis}[
        scale = 1,
        ymode=log,
        xmin=10,xmax=23,
        ymin=7.0E-06,ymax=0.12,
        xlabel={$E_b/N_0$ [dB]}, xlabel style={yshift=0.1em},
        ylabel={BER}, ylabel style={yshift=-0.75em},
        xtick={10,11,12,13,14,15,16,17,18,19,20,21,22,23},
        xticklabels={10,11,12,13,14,15,16,17,18,19,20,21,22,23},
        grid=both,
        ymajorgrids=true,
        xmajorgrids=true,
        grid style=dashed,
        width=0.9\linewidth,
        thick,
        mark size=1,
        legend style={
          nodes={scale=0.75, transform shape},
          anchor={center},
          cells={anchor=west},
          column sep= 1mm,
          row sep= -0.25mm,
          font=\fontsize{6.5pt}{7.2}\selectfont,
        },
        legend columns=1,
        legend pos=south west,
    ]

    \addplot[
        color=myyellow,
        mark=|,
        line width=0.25mm,
        mark size=1.9,
        fill opacity=0,
    ]
    table {
      10  0.102435
      11  0.081746
      12  0.0588663
      13  0.0435333
      14  0.0261488
      15  0.0170428
      16  0.0104507
      17  0.00814653
      18  0.00540291
      19  0.00463812
      20  0.00383163
      21  0.00288359
      22  0.0028404
      23  0.0023359
      24  0.00288321
      25  0.00304107
    };
    \addlegendentry{EBRDF-BP, $d_f=2$ \cite{Jun08_JSAC}}

    \addplot[
        color=myblues,
        mark=diamond*,
        line width=0.25mm,
        mark size=1.9,
        fill opacity=0,
    ]
    table {
      10  0.0639657
      11  0.0501852
      12  0.0381881
      13  0.0268341
      14  0.0168736
      15  0.0110042
      16  0.00704017
      17  0.0040058
      18  0.00214679
      19  0.00116125
      20  0.000466508
      21  0.000231638
      22  0.000102714
      23  4.43979e-05
      24  1.51766e-05
    };
    \addlegendentry{MMSE}


    \addplot[
        color=myred,
        mark=x,
        line width=0.25mm,
        mark size=1.6,
        fill opacity=0,
    ]
    table {
      10  0.0578
      11  0.0466
      12  0.0290
      13  0.0204
      14  0.0125
      15  0.0074
      16  0.0035
      17  0.0024
      18  0.0018
      19  0.0012
      20  0.0010
      21  0.0011
      22  0.0012
      23  0.0011
      24  0.001
      25  0.0011
    };
    \addlegendentry{RD-GAI-BP \cite{yang2018low}}

     \addplot[
        color=mygreen,
        mark=square*,
        line width=0.25mm,
        mark size=1.6,
        fill opacity=0,
    ]
    table {
      10  0.0710034
      11  0.0588184
      12  0.0466429
      13  0.0295864
      14  0.0194492
      15  0.0108398
      16  0.00573001
      17  0.00264177
      18  0.00131724
      19  0.000499913
      20  0.000184289
      21  9.13612e-05
      22  3.71854e-05
      23  1.2778e-05
    };
    \addlegendentry{Proposed BsP with $\mathcal{B}(1,1)$}

    \addplot[
        color=mypurple,
        mark=oplus*,
        line width=0.25mm,
        mark size=1.6,
        fill opacity=0,
    ]
    table {
      10  0.0745823
      11  0.0574884
      12  0.0353821
      13  0.0267995
      14  0.0148194
      15  0.00905567
      16  0.00409888
      17  0.00154513
      18  0.000632859
      19  0.000204275
      20  7.45924e-05
      21  2.59535e-05
    };
    \addlegendentry{Proposed BsP with $\mathcal{B}(2,2)$}

    \addplot[
        color=mypink,
        mark=triangle*,
       line width=0.25mm,
        mark size=1.6,
        fill opacity=0,
    ]
    table {
      10  0.0562
      11  0.0429
      12  0.0257
      13  0.0174
      14  0.0095
      15  0.0050
      16  0.0019
      17  7.6610e-04
      18  2.8756e-04
      19  8.1507e-05
      20  2.5042e-05
      21  7.2917e-06
    };
    \addlegendentry{SD}

    \end{axis}
    \end{tikzpicture} \\
  \caption{BER Performance of the BsP detection and the SOA BP detection algorithms in the MIMO system with $N_r=8$, $N_t=4$ and $64$-QAM modulation.}
  \label{fig:performanceTx4Rx8_64QAM}
\end{figure}

Fig.~\ref{fig:performanceTx4Rx8} reports the BER performance of the BsP detection and the SOA BP detection algorithms in the MIMO system with $N_r=8$, $N_t=4$ and $16$-QAM modulation.
Compared with the MAP detection, the BsP detection with the configuration set $\mathcal{B}(2,2)$ only suffers from performance loss of less than $0.5\,\mathrm{dB}$ at the BER of $10^{-4}$, and the BsP detection with the configuration set $\mathcal{B}(1,1)$ sacrifices $1.7 \; \mathrm{dB}$ performance in the same BER. Compared with the MMSE detection, the BsP detection with the configuration set $\mathcal{B}(1,1)$ earns $1.6\,\mathrm{dB}$ performance gain at the BER of $10^{-4}$, while the BsP detection with the configuration set $\mathcal{B}(2,2)$ achieves $2.8\,\mathrm{dB}$ performance gain at the same BER.
Compared with the RD-GAI-BP detection, the BsP detection with the configuration set $\mathcal{B}(1,1)$ and $\mathcal{B}(2,2)$ reach comparable BER performance in small $E_b/N_0$ regions but avoid performance error floor in high $E_b/N_0$ regions. It should be noted that the Original-BP detection also suffers from performance degradation in high $E_b/N_0$ regions, due to its loopy FG.
Therefore, it is observed that the BsP detection outperforms the SOA BP detection algorithms in high $E_b/N_0$ regions.

\begin{figure}[t]
  \centering
  \begin{tikzpicture}
    \definecolor{myblued}{RGB}{0,114,189}
    \definecolor{myred}{RGB}{217,83,25}
    \definecolor{myyellow}{RGB}{237,137,32}
    \definecolor{mypurple}{RGB}{126,47,142}
    \definecolor{myblues}{RGB}{77,190,238}
    \definecolor{mygreen}{RGB}{32,134,48}
    \definecolor{mypink}{RGB}{255,62,150}
      \pgfplotsset{
        label style = {font=\fontsize{9pt}{7.2}\selectfont},
        tick label style = {font=\fontsize{7pt}{7.2}\selectfont}
      }

    \begin{axis}[
        scale = 1,
        ymode=log,
        xmin=5.0,xmax=17,
        ymin=7.0E-06,ymax=0.2,
        xlabel={$E_b/N_0$ [dB]}, xlabel style={yshift=0.1em},
        ylabel={BER}, ylabel style={yshift=-0.75em},
        xtick={5,6,7,8,9,10,11,12,13,14,15,16,17},
        xticklabels={5,6,7,8,9,10,11,12,13,14,15,16,17},
        grid=both,
        ymajorgrids=true,
        xmajorgrids=true,
        grid style=dashed,
        width=0.9\linewidth,
        thick,
        mark size=1,
        legend style={
          nodes={scale=0.75, transform shape},
          anchor={center},
          cells={anchor=west},
          column sep= 1mm,
          row sep= -0.25mm,
          font=\fontsize{6.5pt}{7.2}\selectfont,
        },
        legend columns=1,
        legend pos=south west,
    ]

    \addplot[
        color=myyellow,
        mark=|,
        line width=0.25mm,
        mark size=1.9,
        fill opacity=0,
    ]
    table {
      5  0.0941276
      6  0.0695353
      7  0.0522483
      8  0.0341005
      9  0.0213536
      10 0.0100526
      11 0.00454382
      12 0.00233922
      13 0.00155853
      14 0.00115787
      15 0.000908097
      16 0.000881425
      17 0.000759985
    };
    \addlegendentry{EBRDF-BP, $d_f=4$ \cite{Jun08_JSAC}}

    \addplot[
        color=myblues,
        mark=diamond*,
        line width=0.25mm,
        mark size=1.9,
        fill opacity=0,
    ]
    table {
      5  0.0845126
      6  0.0640534
      7  0.0504994
      8  0.0389755
      9  0.023857
      10 0.0176028
      11 0.00947273
      12 0.00492542
      13 0.00288964
      14 0.0011644
      15 0.000462016
      16 0.000180423
      17 6.0837e-05
      18 1.466e-05
      19 4.11607e-06
      20 1.03487e-06
    };
    \addlegendentry{MMSE}

    \addplot[
        color=myred,
        mark=x,
       line width=0.25mm,
        mark size=1.9,
        fill opacity=0,
    ]
    table {
      5  0.0847
      6  0.0622
      7  0.0414
      8  0.0279
      9  0.0133
      10 0.0063
      11 0.0031
      12 0.0013
      13 6.1038e-04
      14 3.7722e-04
      15 3.2431e-04
      16 3.2427e-04
      17 3.0051e-04
      18 2.585e-04
      19 2.2720e-04
      20 2.9941e-04
    };
    \addlegendentry{RD-GAI-BP \cite{yang2018low}}


     \addplot[
        color=mygreen,
        mark=otimes*,
        line width=0.25mm,
        mark size=1.6,
        fill opacity=0,
    ]
    table {
      5  0.093564
      6  0.0763504
      7  0.056296
      8  0.0382883
      9  0.0209286
      10 0.0103432
      11 0.00467098
      12 0.00165677
      13 0.000507344
      14 0.000136404
      15 3.80796e-05
      16 1.12553e-05
    };
    \addlegendentry{Proposed BsP with $\mathcal{B}(1,1)$}

    \addplot[
        color=mypurple,
        mark=oplus*,
        line width=0.25mm,
        mark size=1.6,
        fill opacity=0,
    ]
    table {
      5   0.101886
      6   0.0711898
      7   0.0539835
      8   0.0332805
      9   0.0185245
      10  0.00836627
      11  0.00325654
      12  0.00093126
      13  0.00027332
      14  5.87565e-05
      15  1.26639e-05
    };
    \addlegendentry{Proposed BsP with $\mathcal{B}(2,2)$}

    \addplot[
        color=mypink,
        mark=triangle*,
       line width=0.25mm,
        mark size=1.6,
        fill opacity=0,
    ]
    table {
      5  0.0736
      6  0.0548
      7  0.0375
      8  0.0214
      9  0.0109
      10 0.0046
      11 0.0017
      12 6.2360e-04
      13 1.5940e-04
      14 3.7801e-05
      15 7.6250e-06
    };
    \addlegendentry{SD}

    \end{axis}
    \end{tikzpicture} \\
  \caption{BER Performance of the BsP detection and the SOA BP detection algorithms in the MIMO system with $N_r=16$, $N_t=8$ and $16$-QAM modulation.}
  \label{fig:performanceTx8Rx16}
\end{figure}
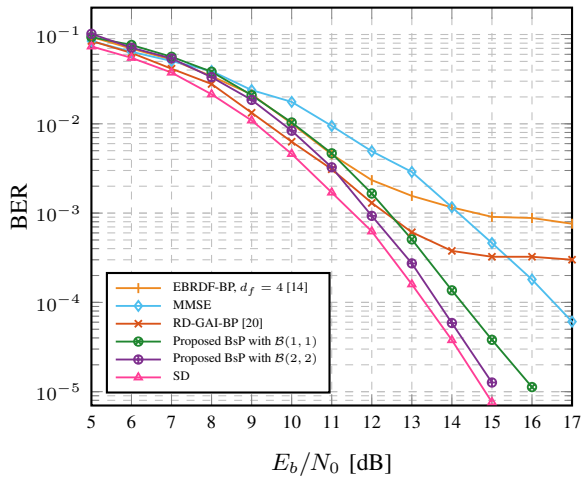

Fig.~\ref{fig:performanceTx4Rx8_64QAM} shows the error rate performance of the BsP detection in the MIMO system with the same antenna scale as that in Fig.~\ref{fig:performanceTx4Rx8}, but with a higher order modulation of $64$-QAM.
The Original-BP detection is not shown here because its complexity is prohibitively high and therefore will not be considered by applications. The MAP detection is also omitted for the same reason. Instead, the SD \cite{Brink03_TCOM} with a large enough radius is simulated to acquire the near-optimal performance.
It can be seen that the BsP detection and the SOA BP detection algorithms exhibit the consistent error rate performance trends as that in Fig.~\ref{fig:performanceTx4Rx8}. Therefore, Fig.~\ref{fig:performanceTx4Rx8_64QAM} demonstrates that the BsP detection can compete further with the SOA BP detection algorithms in MIMO systems with higher order modulations.
Fig.~\ref{fig:performanceTx8Rx16} illustrates the error rate performance of the BsP detection and the SOA BP detection algorithms in the MIMO system with $N_r=16$, $N_t=8$ and $16$-QAM modulation.
The SD detection is the same as that in Fig.~\ref{fig:performanceTx4Rx8_64QAM}.
It is observed that the BsP detection with the configuration set $\mathcal{B}(2,2)$ approaches the SD detection with less than $0.2 \; \mathrm{dB}$ performance loss at the BER of $10^{-4}$. The BsP detection with the configuration set $\mathcal{B}(1,1)$ suffers from $0.95 \; \mathrm{dB}$ performance loss compared with the SD detection at the same BER.
Compared with the RD-GAI-BP detection, the BsP detection with the configuration set $\mathcal{B}(1,1)$ suffers from a slight performance loss in the relatively low $E_b/N_0$ regions, but eliminates the performance error floor in the high $E_b/N_0$ regions.
Fig.~\ref{fig:performanceTx8Rx16} demonstrates that the BsP detection still outperforms the SOA BP detection algorithms in MIMO systems with larger scale antennas.

    \subsection{Performance Comparison for Different Configuration Sets}\label{subsec:differentSets}
\begin{figure}[t]
  \centering
  \begin{tikzpicture}
    \definecolor{myblued}{RGB}{0,114,189}
    \definecolor{myred}{RGB}{217,83,25}
    \definecolor{myyellow}{RGB}{237,137,32}
    \definecolor{mypurple}{RGB}{126,47,142}
    \definecolor{myblues}{RGB}{77,190,238}
    \definecolor{mygreen}{RGB}{32,134,48}
    \definecolor{mypink}{RGB}{255,62,150}
      \pgfplotsset{
        label style = {font=\fontsize{9pt}{7.2}\selectfont},
        tick label style = {font=\fontsize{7pt}{7.2}\selectfont}
      }

    \begin{axis}[
        scale = 1,
        ymode=log,
        xmin=5.0,xmax=17,
        ymin=4.0E-06,ymax=0.2,
        xlabel={$E_b/N_0$ [dB]}, xlabel style={yshift=0.1em},
        ylabel={BER}, ylabel style={yshift=-0.75em},
        xtick={5,6,7,8,9,10,11,12,13,14,15,16,17},
        xticklabels={5,6,7,8,9,10,11,12,13,14,15,16,17},
        grid=both,
        ymajorgrids=true,
        xmajorgrids=true,
        grid style=dashed,
        width=0.9\linewidth,
        thick,
        mark size=1,
        legend style={
          nodes={scale=0.75, transform shape},
          anchor={center},
          cells={anchor=west},
          column sep= 1mm,
          row sep= -0.25mm,
          font=\fontsize{6.5pt}{7.2}\selectfont,
        },
        legend columns=1,
        legend pos=south west,
    ]

     \addplot[
        color=mygreen,
        mark=otimes*,
        line width=0.25mm,
        mark size=1.6,
        fill opacity=0,
    ]
    table {
      5  0.104432
      6  0.0765957
      7  0.0518771
      8  0.0444236
      9  0.0254859
      10 0.0148458
      11 0.00853277
      12 0.00425407
      13 0.00201504
      14 0.000845318
      15 0.000346684
      16 0.000131103
      17 5.0784e-05
      18 1.73276e-05
      19 7.34441e-06

    };
    \addlegendentry{Proposed BsP with $\mathcal{B}(1,1)$}

    \addplot[
        color=mypurple,
        mark=oplus*,
        line width=0.25mm,
        mark size=1.6,
        fill opacity=0,
    ]
    table {
      5   0.0844933
      6   0.0681497
      7   0.0573657
      8   0.0346091
      9   0.0218156
      10  0.0114301
      11  0.00574092
      12  0.00256042
      13  0.00095799
      14  0.000347315
      15  0.000117874
      16  3.36968e-05
      17  1.07373e-05
    };
    \addlegendentry{Proposed BsP with $\mathcal{B}(2,2)$}

    \addplot[
        color=myyellow,
        mark=|,
        line width=0.25mm,
        mark size=1.9,
        fill opacity=0,
    ]
    table {
      5  0.0879425
      6  0.0721154
      7  0.0543561
      8  0.0356859
      9  0.0219867
      10 0.0117982
      11 0.00595153
      12 0.00285739
      13 0.000958332
      14 0.000335268
      15 0.00010727
      16 3.60753e-05
      17 1.15644e-05
    };
    \addlegendentry{Proposed BsP with $\mathcal{B}(8,2)$}

    \addplot[
        color=myblues,
        mark=diamond*,
        line width=0.25mm,
        mark size=1.9,
        fill opacity=0,
    ]
    table {
      5  0.0880012
      6  0.0694972
      7  0.0512845
      8  0.0350297
      9  0.0228076
      10 0.0109489
      11 0.0057111
      12 0.00212679
      13 0.000873694
      14 0.000355324
      15 9.94445e-05
      16 2.92991e-05
      17 9.69366e-06
      18 2.71855e-06
    };
    \addlegendentry{Proposed BsP with $\mathcal{B}(3,3)$}

    \addplot[
        color=myred,
        mark=x,
       line width=0.25mm,
        mark size=1.9,
        fill opacity=0,
    ]
    table {
      5  0.09375
      6  0.0700188
      7  0.0477086
      8  0.0320912
      9  0.0188189
      10 0.0113636
      11 0.00528054
      12 0.00242707
      13 0.00094263
      14 0.000327696
      15 0.000117843
      16 3.17958e-05
      17 9.85711e-06
    };
    \addlegendentry{Proposed BsP with $\mathcal{B}(4,4)$}

    \addplot[
        color=mypurple,
        mark=*,
        line width=0.25mm,
        mark size=1.6,
        fill opacity=0,
    ]
    table {
      5   0.090483
      6   0.0678267
      7   0.0487259
      8   0.0349124
      9   0.0216619
      10  0.0119144
      11  0.0059956
      12  0.0023843
      13  0.000875671
      14  0.000301414
      15  9.91736e-05
      16  3.79607e-05
    };
    \addlegendentry{Proposed BsP with $\mathcal{B}(8,4)$}

    \addplot[
        color=mypink,
        mark=triangle*,
       line width=0.25mm,
        mark size=1.6,
        fill opacity=0,
    ]
    table {
      5  0.0755965
      6  0.0585664
      7  0.0391406
      8  0.02358
      9  0.0146888
      10 0.00764489
      11 0.00334853
      12 0.00137913
      13 0.000586802
      14 0.000195404
      15 6.03787e-05
      16 1.57112e-05
      17 4.4188e-06
    };
    \addlegendentry{MAP}

    \end{axis}
    \end{tikzpicture}
  \caption{BER performance for the BsP with different configuration sets in the
MIMO system with $N_r = 8$, $N_t = 4$ and $16$-QAM modulation.}
  \label{fig:performanceTx4Rx8Config}
\end{figure}
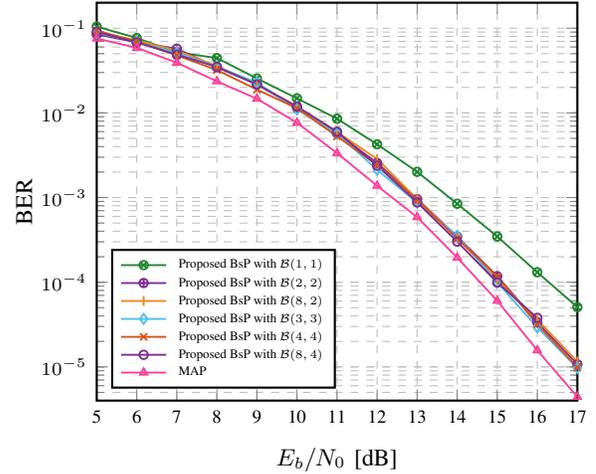
The BER performance of the BsP detection with different configuration sets are compared in this subsection. Empirically, the BsP detection with the configuration set involving a large $d_m$ and $d_f$ will achieve a better BER performance, due to its large search space of the transmitted symbol vector.
However, Fig.~\ref{fig:performanceTx4Rx8Config} exhibits different experimental results in the MIMO system with $N_r = 8$, $N_t = 4$ and $16$-QAM modulation.
It is observed that the configuration set $\mathcal{B}(2,2)$ contributes to $1.2 \; \mathrm{dB}$ performance gain for the BsP detection at the BER of $10^{-4}$, compared with the configuration set $\mathcal{B}(1,1)$. This performance gain is consistent with the empirical recognition. However, compared with the configuration set $\mathcal{B}(2,2)$, the configuration sets with large $d_m$ and $d_f$ fail in further bringing performance gains to the BsP detection.
The reason is straightforward: Since the MMSE detector provides an effective initial \textit{a priori} information for each transmitted symbol in high $E_b/N_0$ regions, the configuration set with a large $d_m$ and $d_f$ has a very small probability to contain the optimal transmitted symbol vector in its additional search space compared with the configuration set $\mathcal{B}(2,2)$, thus resulting in a limited performance gain.
	
	\section{Conclusions}\label{sec:conclusion}
In this paper, the BsP MIMO detector is proposed to balance performance and complexity. By exploiting the symbol probabilities to truncate the search space of the symbol candidates, the BsP detector enjoys a lower complexity than the SOA BP detectors. This approach also helps to modify the loopy structure of the FG and eliminate the performance error floor. Numerical results have shown that with the proposed simplifying mechanisms, the BsP detector outperforms its counterparts in terms of both performance and complexity. It should be noted that the proposed BsP detector is suitable for message-passing problems with large candidate set, including but not limited to MIMO detection. Future work will be directed towards joint iterative cooperation with other baseband modules, the efficient hardware implementations of the BsP MIMO detector, and other applications of BsP in a generalized fashion.


    \balance

	\bibliographystyle{IEEEtran}
	\bibliography{./bib/IEEEabrv,./bib/IEEEBib}
	
	\balance
	
\end{document}